\documentclass[conference]{IEEEtran}
\usepackage[ruled]{algorithm2e} 
\usepackage{listings, xcolor} 
\usepackage{graphicx} 
\title{DFS: A Dataset File System for Data Discovering Users}
\author{
\IEEEauthorblockN{Yasith Jayawardana and Sampath Jayarathna\\
\texttt{\{yasith, sampath\}@cs.odu.edu}}
\IEEEauthorblockA{Department of Computer Science\\
Old Dominion University\\
Norfolk, VA 23529 USA}
}
\definecolor{background}{HTML}{F0F0F0}
\begin{document}
\maketitle
\begin{abstract}
Many research questions can be answered quickly and efficiently using data already collected for previous research.
This practice is called secondary data analysis (SDA), and has gained popularity due to lower costs and improved research efficiency.
In this paper we propose DFS, a file system to standardize the metadata representation of datasets, and DDU, a scalable architecture based on DFS for semi-automated metadata generation and data recommendation on the cloud.
We discuss how DFS and DDU lays groundwork for automatic dataset aggregation, how it integrates with existing data wrangling and machine learning tools, and explores their implications on datasets stored in digital libraries.
\end{abstract}
\section{Introduction}
With the advancements in digital technology, researchers have access to a vast amount of data collected for past studies.
They are utilized by many research communities to fuel entirely new research or to expand on the original study.
This practice, termed secondary data analysis (SDA), enables conducting non-experimental research with minimal cost.

Due to the nature of the WWW, not all datasets are regulated by an authority, nor follow any universally agreed convention.
As a result, datasets have inconsistent naming conventions and file formats, making it challenging to understand their semantics.
Selecting a dataset for SDA has thus become a complex process that involves searching for datasets, analyzing candidate datasets for applicability, and data wrangling \cite{kandel2011research}.
The required pre-processing varies across different file types and data, and cannot be pre-determined without understanding the nature of data.
This exerts a heavy workload on users to discover and pre-process data, which consumes time and effort that could be utilized more productively at the presence of a unified semantic representation for datasets.

Another challenge in SDA is ensuring the authenticity of datasets. While cryptography plays a major role in ensuring trust and authenticity of digital content, ensuring the authenticity of datasets has not been explored.
Datasets could be easily forged and uploaded to data sharing platforms, and researchers depending on such falsified data could arrive at misleading conclusions in their publications.
These errors cannot be cross-referenced back to the data source without proper citation practices.
Hence datasets require a mechanism to ensure trust and reference immutability, neither of which is available at present.

Under such constraints, the SDA task could become overly complex, which is detrimental to the quality and efficiency of research.
\section{Background}
At present, the approach for discovery of data is to a large extent based on "Users Discovering Data" (UDD).
However, its opposite, "Data Discovering Users (DDU)" is now gaining traction with the introduction of dataset repositories, search platforms and recommender systems.
Commercial retailers increasingly use advanced algorithms including big data analytics, deep learning, deep search, and crowd-sourcing to enable such interactions.
Thus, theoretical concepts have been developed to capture connections between agents, products, tools, activities and transactions, and to construct graph data describing the chains and networks between these elements.
\subsection{Related Work}
There has been recent studies attempting to link open data sets based on the presence of schema overlap between datasets \cite{ellefi2016dataset}.
Each linked datasets on the Linked Open Data Cloud were characterized through a set of schema concept labels that describe them.
Schema overlaps were identified using a semantico-frequential concept similarity measure and a ranking criterium based on TF-IDF cosine similarity.
The mappings between the schema concepts were also obtained.
Through this, they have obtained an average precision of up to 53\% for a recall of 100\%.

Another study \cite{10.1007/978-3-319-07443-6_35} proposes a mechanism to create linked dataset profiles.
Each profile consists of structured dataset metadata describing topics and their relevance, all generated by sampling resources from datasets, extracting topics from reference datasets, and ranking based on graphical models.
They have created topic profiles for all accessible datasets in the Linked Open Data Cloud and showed that this approach generates accurate profiles even with comparably small sample sizes (10\%), while outperforming established topic modelling approaches.

Parekh \cite{DBLP:conf/ike/ParekhGF04} has proposed an ontology-based semantic metadata paradigm using Semantic Web languages.
They have defined elements to incorporate information about data identification, spatial extent, temporal extent, data presentation form, data content and data distribution regarding the dataset, and allowed data providers to select concepts from domain ontologies that best describe the dataset.
These selections, along with links to domain ontologies were stored in a metadata file, thereby generating semantic metadata for datasets that facilitate content based discovery of datasets irrespective of their locations and formats.

Another study \cite{bhattacherjee2015principles} provides a storage-efficient approach to provide version control to datasets.
They state that the amount of storage used is proportional to the speed of recreating or retrieving dataset versions.
A suite of inexpensive heuristics were created based on techniques in delay-constrained scheduling, and spanning tree literature.
Results show that these heuristics provide efficient solutions in practical dataset versioning scenarios.
\subsection{Secondary Data Analysis (SDA)}
A typical SDA task is conducted in several stages.
Initially, data needs to be discovered using dataset search or obtained directly from collaborators.
The datasets may come from different sources, and should be aggregated for analysis.
Next, the data should be loaded into a data manipulation/visualization tool to determine their relevance and identify any preprocessing needed.
Following this, the data should be preprocessed.
Here, redundant data should be cleaned, and complementary data should be aggregated through identification of matching fields and patterns.
If multiple datasets are used, the user needs to determine how to join them into a single data pool.
Finally, the user should drop the attributes irrelevant to the hypothesis, and apply any algorithms needed to model their hypotheses to obtain results.
\subsection{Data Wrangling}
Data Wrangler \cite{kandel2011wrangler} is a tool developed by researchers at Stanford to simplify the task of "data wrangling", which involves reformatting data values or layout, correcting erroneous or missing values, and integrating multiple data sources.
It attempts to automatically infer the required transforms for cleaning and organizing the data, and leverages semantic data types (e.g., geographic locations, dates, classification codes) to aid validation and type conversion.
These semantic data types are probabilistic estimates from the provided data, and is prone to errors.
Thus, having rich and accurate semantic data is vital for wrangling the data successfully.
In the presence of such semantic data, Data Wrangler can potentially infer the transforms with higher confidence, and apply them automatically to prepare datasets for studies.
\subsection{Automated Machine Learning}
There is a research trend towards automatic machine learning, which has led to the development of frameworks such as Auto-Keras \cite{jin2018efficient}, and Data Wrangler  \cite{kandel2011wrangler}.
Auto-Keras is a Neural Architecture Search (NAS) to select an optimal configuration for training a neural network for the given data based on Keras \cite{gulli2017deep}.
Research efforts spanning across many domains have been made using NAS for automating machine learning \cite{zoph2016neural, liu2018progressive, tan2018mnasnet}.
If the dataset semantics are readily available, it could act as a heuristic for optimizing the NAS process in Auto-Keras and simplify the learning process further.
\subsection{File Formats and Metadata}
Different data types are represented using in different file formats specifically optimized for them. CSV/XLSX targets tabular data, PNG/MP3/MP4 targets multimedia, and PDF/DOCX/HTML targets documents.
Formats such as CDF/GRIB \cite{luengo2004grid} targets storage efficiency, while formats such as RDF \cite{quilitz2008querying} provides the ability to store semantic relationships.
However, not all file formats can store metadata.
Hence, a file system that maintains metadata and links related files together is ideal for supporting a wide variety of datasets.

HDF5 \cite{folk2011hdf5} is a file container that supports metadata.
It provides file archival, and maintains hierarchical information along with metadata.
However, transmitting gigabytes of data across a network just to compare metadata is not an efficient solution.
To support data recommendation, aggregation and clustering while preserving the immutability of data, metadata should be maintained external to the data files.
Studies \cite{brimhall2001removing} show that researchers without a computer science background prefers integrated single-click graphical programs over terminal and code.
Thus, having an efficient metadata format will help in generating text summaries and creating graphical programs that works for any type of dataset, and in turn, improve cross-domain research productivity.
\section{Hypothesis}
We hypothesize that a standardized metadata format would compensate for the tedious preprocessing and knowledge gathering steps required to understand and interpret poorly documented datasets, by streamlining information management in datasets, introducing dataset versioning, and laying the groundwork for rule-based and machine learning algorithms to generate metadata and data recommendations.
Data recommendations provided based on user interest would connect like-minded users, and increase collaboration on datasets in digital libraries.
This revolutionizes the current approach to data discovery, sharing and usage, and by doing so greatly enhance the exploitation of available open-access data sets for research and the realization of societal benefits.
\section{Key Concepts}
The subsections below introduce the key components of the envisioned architecture for managing datasets in digital libraries.
\subsection{Data Discovering Users (DDU)}
\begin{figure}[hbt!]
  \centering
  \includegraphics[width=\linewidth]{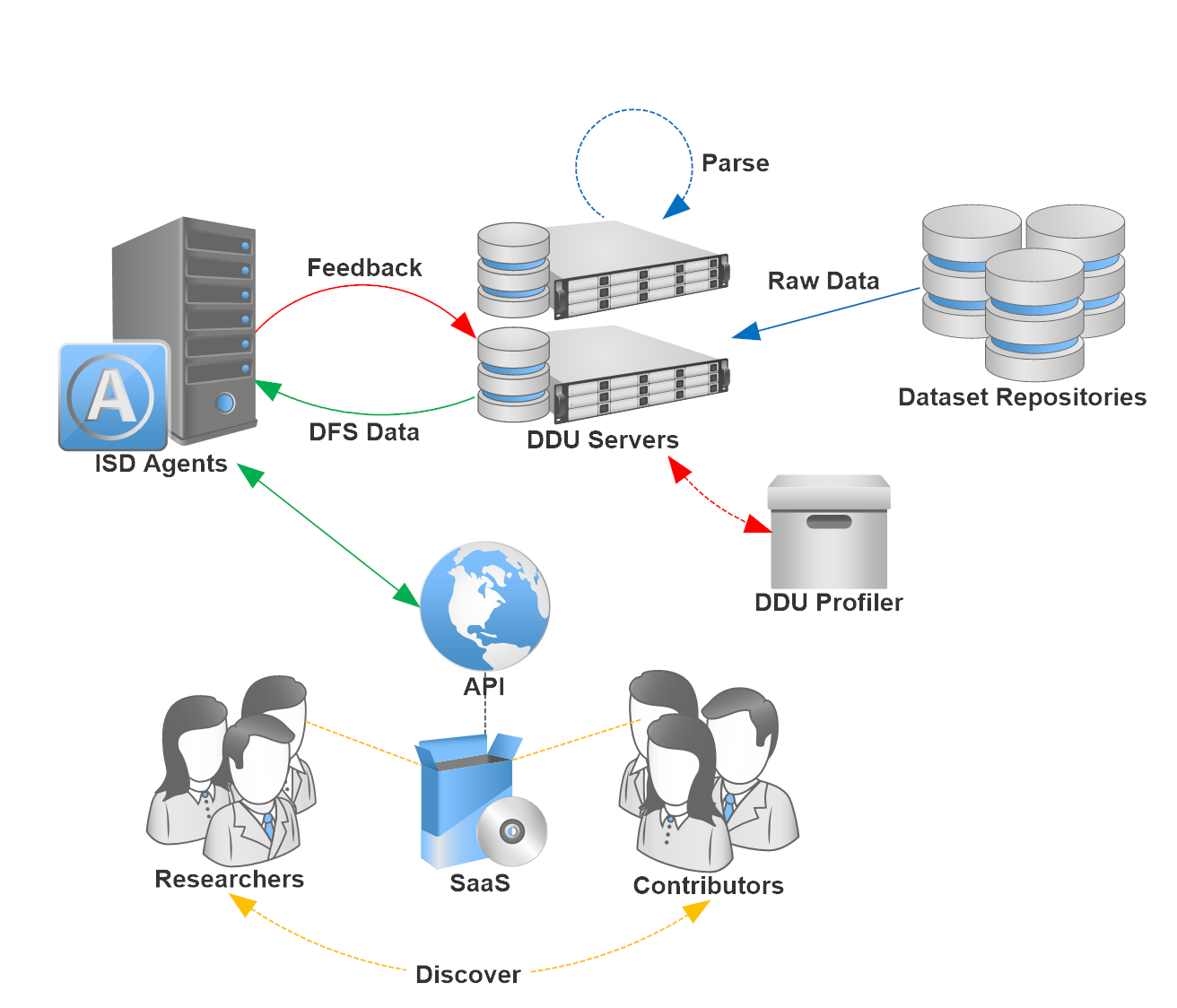}
  \caption{Architecture of DDU including DDU Profilers \cite{jayarathna2017analysis}}
  \label{fig:architecture}
\end{figure}

In DDU, datasets and products derived from interactions will be associated with "Intelligent Semantic Data" (ISD) that can communicate semantic information in response to queries including access conditions, quality, uncertainties, guidance on applicability, and user feedback.
Figure \ref{fig:architecture} shows an overview of how different components are interconnected in DDU.
Datasets are stored in repositories, which are responsible for handling data replication and version control.
The DDU servers maintain metadata files (i.e. metafiles) for each dataset, and indexes datasets by their fields.
The metadata is generated using machine learning and improved through crowd-sourcing.
The ISD Agents expose APIs for users to query, discover and fetch datasets.
Crowd-sourced metadata is fed into the ISD Agents, which in turn, improves the existing metadata on DDU Servers.
DDU Profilers maintain interest profile models (IPMs) \cite{jayarathna2017analysis} for tracking user interests and providing intelligent matches.
The user-facing component, which is DDU SaaS, utilizes all aforementioned components to provide an ecosystem for collaborative research.
\subsection{Dataset File System}
The main component of DFS is the metadata file (or metafile), which serves as the entry point to a dataset.
Its objective is to capture as much information as possible about the underlying dataset, such that it eliminates the need to rely on external documentation to understand the dataset semantics.
Each metafile stores information about the dataset, data files, and data fields.
This enables multiple data files to behave as one coherent set of data, and also allows data files to be shared across datasets.

Figure \ref{fig:dfs-architecture} shows a sample metafile (shortened for brevity) in DFS.
\begin{figure}[hbt!]
\begin{lstlisting}[firstline=1,lastline=51]
{
    "$schema": "https://some.example.com/schema-v1.3.json",
    "$id": "A2BE-FC28-906C-03B7",
    "meta-version": 2,
    "created": "01-20-2019",
    "modified": "05-27-2019",
    "checksum": 43278947328957439805847390257439,
    "meta": {
        "name": "EEG recordings during ADOS-2",
        "description": "...",
        "copyright": ".....",
        "keywords": ["EEG", "ADOS-2", "ASD", "MEDICINE"],
        "authors": [{
            "$id": "023-23-425325",
            "name": "John Doe",
            "affiliation": "Some University",
            "email" : "john@example.edu"
        }],
        "files": [{
            "$id": 21,
            "path": "./020.json",
            "encoding": "JSON",
            "version": 1,
            "checksum": 0123015035783941274895378,
            "description": "Participant 020",
            "measurement": {
                "name": "Brain Activity",
                "device": "Emotiv EPOC+",
                "units": "mV"
            },
            "fields": [{
                "name": "Fpz",
                "type": "Float", 
                "description": "Electrode Fpz"
            },
            {
                "name": "Oz",
                "type": "Float", 
                "description": "Electrode Oz"
            }]
        }],
        "links": [{
            "$type": "ID",
            "description": "Subject ID",
            "fields": [
                {"file_id": 1, "field":"subject"},
                {"file_id": 2, "field":"*"}
            ]
        }]
    }
}
\end{lstlisting}
    \caption{Sample Metafile in DFS (Shortened for Brevity)}
    \label{fig:dfs-architecture}
\end{figure}
Each metafile contains a "\$schema" field to identify the JSON schema, and the fields "id" and "meta-version" to uniquely identify a dataset.
The "meta-version" increments as the metadata and data files mature over time.
The "created", and "modified" fields provide a timeline of any changes to the data or metadata.
The "checksum" field maintains the hash of the contents in the "meta" field, which enforces integrity.

The "meta" field keeps all information related to the dataset.
The "name" and "description" fields describe the dataset in a human-readable format, and the "copyright" field stores any copyright information.
The "keywords" field indicate the research sub-domains, and is updated dynamically through crowd-sourcing and user profiling.
The "author", field maintains a list of dataset authors, and provides fields to store their "\$id", "name", "affiliation" and "email" for authenticity.

The metafile points to data files through "files" field.
The "files" field maintains an "\$id" field for internal reference.
Its "path" and "encoding" fields store the file path and format to read the file.
The "version" field increments each time the file is changed, and the "checksum" field provides integrity by ensuring that files cannot be arbitrarily modified without invalidating the metadata.
The "description" field provides a textual description of the file, and the "measurement" provide the type of measurement, the devices used to measure it, and the units the measurements are stored in the file.
The "fields" field keeps a list of each field in the file, and includes their type and description to provide better semantics.

Following the "files" field, the "links" field keeps track of the semantic relationships between the fields in data files. For example, if the Field X of File A maps to Field Y of File B, this can be stored as a link by providing the link type (e.g. ID), a description of the link, and the fields involved in the link.

As an added benefit, the "id" and "meta-version" fields provide version control capability and reference immutability, making them an ideal candidate for citation.
Citing datasets using DFS would resolve ambiguities caused by evolving datasets.
\section{Applications of DFS}
Introducing DFS creates a significant impact on existing and future research. 
It provides the infrastructure for representing dataset semantics in rich detail, effectively eliminating dependencies on external sources to comprehend them.

The most obvious application of DFS is for DDU.
Here, DFS lays the groundwork for semi-automated metadata management in DDU.
Apart from DDU, we identified several applications of DFS that are described in the sections below.
\subsection{SDA Automation}
For the data acquisition stage of SDA, DFS provides the groundwork for semantic searching, and could potentially provide targeted results based on the domain and topic of research.
DDU realizes this concept by including user interest as a factor for dataset search.
For the preprocessing stage, the metafiles serve as a semantic marker to determine the preprocessing needed for each data file in a dataset.
Transforms could be inferred semantic information in the metafiles, and applied on the source data to obtain the target format.
Complementary data could be aggregated, and redundant data could be removed during this process by identifying the semantic similarities between each field.
Section \ref{dataset_aggregation} discusses how data could be aggregated using DFS metadata in more detail.
When evaluating the hypotheses, the semantic information could be leveraged by machine learning libraries to identify the type of model required. If neural networks are involved, the semantic information can be used to heuristically tune the model hyper-parameters using NAS.
\subsection{Integration With Existing Tools}
\begin{figure}[hbt!]
  \centering
  \includegraphics[width=\linewidth]{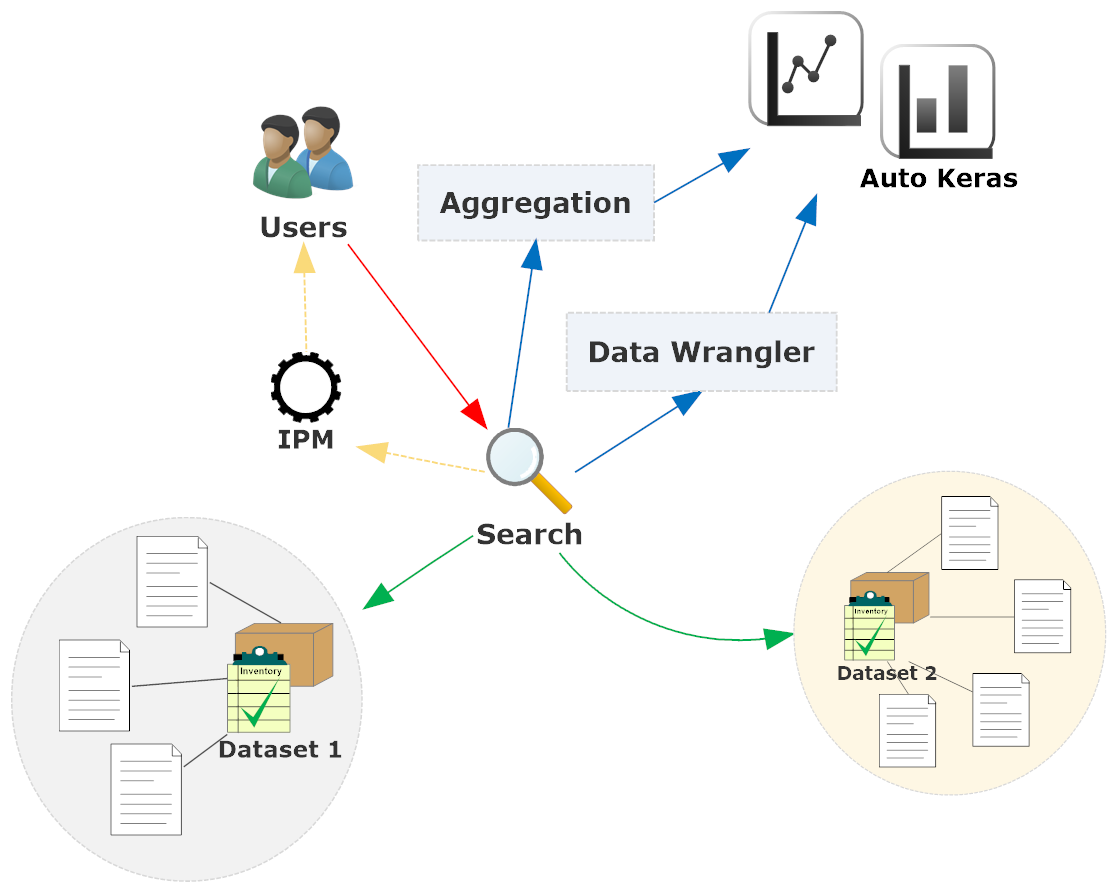}
  \caption{Integration with Data Wrangler and Auto-Keras}
  \label{fig:dfs-2}
\end{figure}

Figure \ref{fig:dfs-2} shows how DDU and DFS can be complemented with existing tools and platforms to streamline data analytics.
Here, the user queries can be cross-referenced with IPM \cite{jayarathna2017analysis} to conducts user profiling and tune the results.
Data Wrangler \cite{kandel2011wrangler} can leverage the semantic information in DFS to generate the transforms required to pre-processes data.
Data aggregation, as described in Section \ref{dataset_aggregation}, can be used to multiple datasets based on relevance, and enables automatic clustering of data.
The resulting aggregate data and metadata can then be passed to Auto-Keras \cite{jin2018efficient} to heuristically determine the best way to model the hypothesis of the experiments, effectively producing a fully automated pipeline for data analytics with minimal user intervention.
\subsection{Dataset aggregation} \label{dataset_aggregation}
Dataset aggregation is the process of comparing datasets using their field information to determine if they could be merged.
Studies have proposed methods for calculating dataset similarity using schema overlap \cite{ellefi2016dataset}, and for merging datasets using scalable algorithms \cite{yang2009map}.
Since DFS metafiles provide information about the data fields and how they are related to each other, datasets could be compared using their metafiles to determine if they could be merged.

Algorithm \ref{alg:aggregate} provides a pseudo-code for dataset aggregation using DFS.
\begin{algorithm}[tph]
\caption{Dataset Aggregation using Metafiles}
\label{alg:aggregate}
\SetKwProg{aggregate}{function \emph{aggregate}}{ :}{end}
\SetAlgoNoEnd
\aggregate{($\alpha, \beta$)}{
  \If{$similarity(graph(\alpha), graph(\beta)) \leq \epsilon$}{
    \textbf{throw} error\;
  }
  \ForAll{$\gamma \gets fields(\alpha)$}{
    \ForAll{$\delta \gets fields(\beta)$}{
      \If{$overlap(\gamma, \delta) \geq \sigma$}{
        $\alpha \gets metajoin(\alpha, \beta, \gamma, \delta)$\;
      }
    }
  }
  \Return{$\alpha$};
}
\end{algorithm}
For each metafile, the fields and their links are represented as a graph.
Next, the two graphs are compared using graph similarity algorithms to determine if the datasets are comparable.
If so, a join operation is performed on the fields and links of the two metafiles based on the schema overlap.
This results in a connected graph which represents links among both datasets.
This information is then used to create a new metafile which represents data from both datasets.

\begin{figure}[hbt!]
  \centering
  \includegraphics[width=\linewidth]{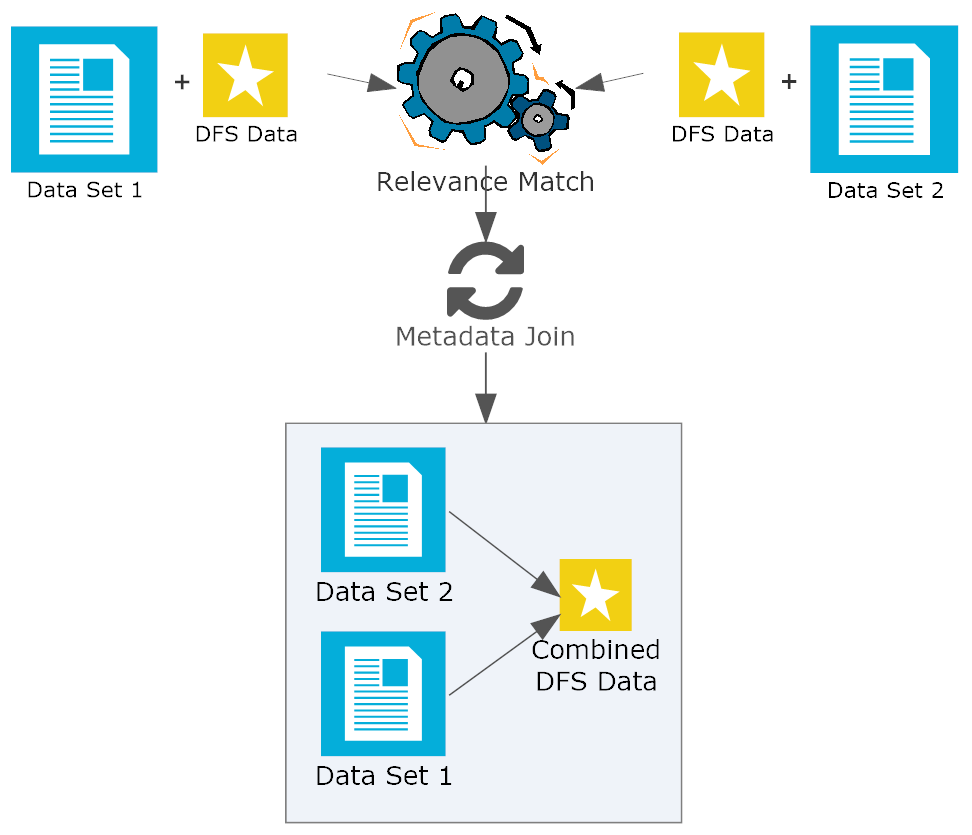}
  \caption{Dataset Aggregation Process}
  \label{fig:dfs-mindmap}
\end{figure}

Figure \ref{fig:dfs-mindmap} provides a visualization of this aggregation process.
Each data set contains meta data that describes the fields (columns in tabular data), domain, tags, encoding, authors, mode of data extraction, etc. that defines the data set in a structured, standardized format. Comparing the meta data of each file, the aggregation process combines the two data sets into a new data set with aggregated metadata.
\section{Conclusion and Future Work}
DFS and DDU provide a fresh outlook to how data is discovered, wrangled, and used for data analytics and machine learning.
With DFS bringing new techniques for dataset aggregation and DDU enabling semi-automated metadata management and user interest profiling, research communities could collaborate efficiently on research and accelerate workflows.
As a future work, we plan to conduct a comprehensive survey with researchers, data curators, and practitioners, and incorporate their feedback to fine tune the DFS schema and the DDU architecture. We also plan to evaluate the compatibility of DFS across multiple domains and file types through scenarios and case studies, to evaluate the cross-domain coverage of DFS.
\bibliographystyle{IEEEtran}

\end{document}